\documentclass{eas}
\usepackage{graphicx}
\newcommand{\myr}{\,$M_{\odot}\,{\rm yr}^{-1}$}

\TitreGlobal{The Physics of Evolved Stars}
\begin{document}

%%-----------------------------
%%      the top matter
%%-----------------------------
\title{Indication of the high mass-transfer ratio 
       in~S-type symbiotic binaries} 
\author{N. Shagatova}
\address{Astronomical Institute of the Slovak Academy 
         of Sciences, Tatransk\'a Lomnica, Slovakia}
\author{A. Skopal}
\sameaddress{1}
%\author{...}\address{...}
\runningtitle{Shagatova \& Skopal: 
              Mass-transfer ratio in S-type symbiotics}
\thanks{
Supported by a grant of the Slovak Academy of Sciences, VEGA No.~2/0002/13.}
\begin{abstract}
By modelling H$^0$ column densities in eclipsing S-type 
symbiotic stars EG And and SY Mus, we derived the wind 
velocity profile and the corresponding mass-loss rate from 
their giants. Our analysis revealed a strong enhancement of 
the wind at the orbital plane. 
%that indicates a high mass-transfer ratio in S-type symbiotic stars. 
%------------------------------------------------------------------
%In this work we determined the velocity profile of the wind 
%from the giants in eclipsing S-type symbiotic binaries EG~And 
%and SY~Mus at the near-orbital-plane region, derived the
%spherical equivalents of the mass-loss rate $\dot{M}_{\rm sp}$ 
%and explored possible implications.  
\end{abstract}
\maketitle
%%-----------------------------
%%      your text
%%-----------------------------
\section{Introduction}

The wind mass transfer in symbiotic binaries is connected 
with the long-standing problem of a large luminosity 
of their hot components and an inefficient wind mass 
transfer from their red giants in the canonical Bondi-Hoyle 
accretion mechanism. 
Investigation of the giant wind properties can aid us in 
a better understanding of the wind mass-transfer mode in these 
widest interacting binaries. 

\section{Method and results}

We investigated a distribution of the neutral hydrogen (H$^0$) 
from the wind of giants in eclipsing S-type symbiotic stars 
EG And and SY Mus. For this purpose, we used far-UV spectra 
measured by the IUE and HST satellites, available from their 
archives. By modelling the Rayleigh attenuation of the continuum 
around the Ly-$\alpha$ line (Fig. 1, left), we obtained 
%(see Nussbaumer \etal\ 1989), 
H$^0$ column densities, $n_{\rm H^0}^{\rm obs}$, at different 
orbital phases (Fig. 1, right). Some values were supplemented 
from the literature. 
Further, we modelled $n_{\rm H^0}^{\rm obs}$ values taking 
into account ionization of the giant's wind (Seaquist \etal\ 1984) 
and used the inversion method for the column density function 
according to Knill~\etal\ (\cite{kn93}) to derive the wind 
velocity profile (WVP). 
\begin{figure}
\begin{center}
 \includegraphics[width=4.5cm,angle=0]{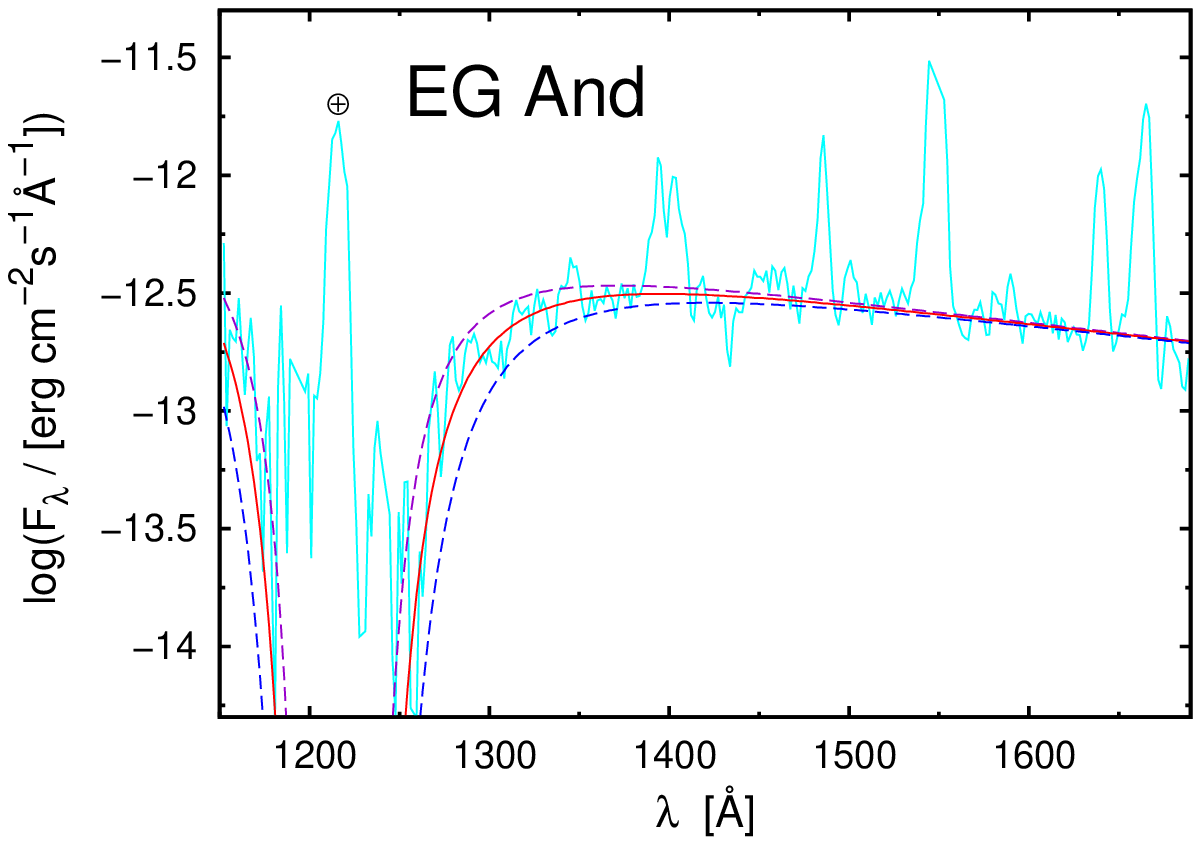}
 \qquad
 \includegraphics[width=4.4cm,angle=0]{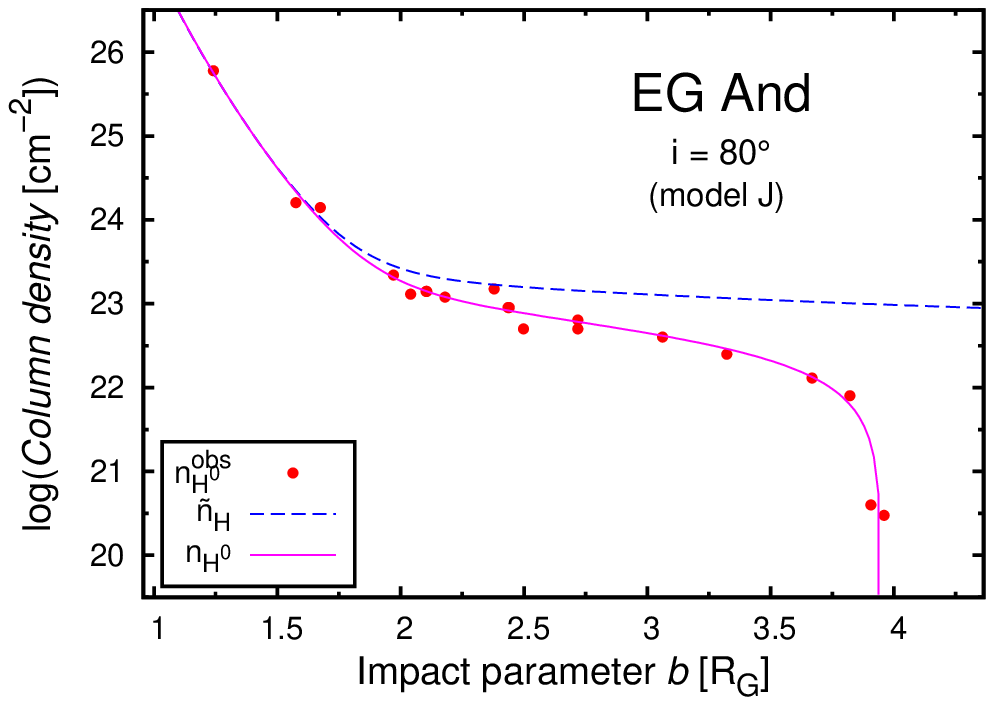}
\end{center}
\vspace*{-5mm}
\caption{Left: Far-UV spectrum of EG~And attenuated with 
$n^{\rm obs}_{\rm H^0} = (1.5 +0.7/-0.5)\times 10^{23}$cm$^{-2}$. 
Right: $n^{\rm obs}_{\rm H^0}$ values (circles) and our models 
(solid and dashed line).
        } 
\end{figure}
In this way we derived a relation for the total H$^0$ column 
density, $\tilde{n}_{\rm H}(b)$, as a function of the orbital 
phase (or the impact parameter $b$), and the WVP, $v(r)$, in 
the form, 
\begin{equation}
\label{rovn}
   \tilde{n}_{\rm H}(b) = \displaystyle\frac{n_1}{b} + 
          \displaystyle\frac{n_K}{b^K}, \hspace{1.5cm}
   v(r) = \displaystyle\frac{v_\infty}{1 + \xi r^{1-K}},
\end{equation}
where $n_1$, $n_{\rm K}$ and $K$ are fitting parameters, 
$v_\infty$ is the terminal velocity of the wind, 
%  $\xi = \displaystyle\frac{n_K\lambda_1}{n_1\lambda_K}$, 
   $\xi = (n_K\lambda_1)/(n_1\lambda_{\rm K})$, 
$\lambda_1$ and $\lambda_{\rm K}$ are the eigenvalues of the Abel 
operator (see Knill \etal\ \cite{kn93}). 
Example of a model and its parameters for EG And are in 
Fig. 1 and Table 1. Corresponding values of the spherical 
equivalent of the mass-loss rates, 
$\dot{M}_{\rm sp} \approx 10^{-6}$\myr, are a factor of 
$\approx 10$ larger than total rates measured by independent 
methods ($\approx 10^{-7}$\myr, e.g. Seaquist \etal\ \cite{s+93}). 
This findings suggests that the wind from giants in S-type 
symbiotic stars is focused towards the binary orbital plane, 
because both systems are eclipsing, and thus our 
$n_{\rm H^0}^{\rm obs}$ values are given by densities at 
the near-orbital-plane region. According to the model of 
Nagae \etal\ (2004), our WVPs correspond to the mass-accretion
ratio $15-18\%$.
%
%------------------------ Table 2 -------------------------
%
\begin{table*}[h!]
\caption{Resulting parameters, $n_1$, $n_{\rm K}$, $K$ and 
         $X^{\rm H+}$; $X^{\rm H+}$ is
         the ionization parameter. 
}
\label{vysl}
\centering
\begin{tabular}{lcccccc}
\hline\hline
object & $i$ & $n_1[10^{23}]$  & $n_{\rm K}$ & $K$ & 
$X^{\rm H+}$ & $\dot{M}_{\rm sp}$ [$M_\odot$yr$^{-1}$] \\
    \hline
EG~And
& $80^{\circ}$ & $3.87$ & $1.15\times 10^{27}$ & $14$ & 
  $1.85$ & $1.8\times 10^{-6}$  \\
%SY~Mus 
%& $84^{\circ}$ & E & $6.20$ & $5.30\times 10^{26}$ & $8 $ &
%  $2.94\times 10^{3}$ & $2.50$   \\
%& $84^{\circ}$ & I & $2.45$ & $1.00\times 10^{27}$ & $13$ & 
%  $1.81\times 10^{4}$ & $16.0$   \\
\hline
%\multicolumn{8}{l}{Notes: $^{1)}$ E -- egress data, I -- ingress data}
\end{tabular}
\end{table*}

\section{Conclusions}

By modelling the H$^0$ column densities around giants in 
eclipsing S-type 
symbiotic stars EG And and SY Mus, we determined the WVP and 
the corresponding $\dot{M}_{\rm sp}$ from their giants. 
Our analysis revealed that $\dot{M}_{\rm sp}$ are a factor 
of $\approx 10$ higher than observed total $\dot{M}$. 
This suggests that the giant's wind is significantly enhanced 
at the orbital plane, where can be effectively accreated onto 
the hot component. 
In this way, we indicated a high mass-transfer ratio in 
S-type symbiotic binaries. 

%
%%-----------------------------
%%      your bibliography
%%-----------------------------

\end{document}